\documentclass[12pt]{article}

\usepackage{graphicx,amsmath,amssymb,url,enumerate,mathrsfs,epsfig,color}
\usepackage{tikz}

\usepackage{pgfplots}
\usepackage{stmaryrd}
\usetikzlibrary{calc, patterns,arrows, shapes.geometric}
\pgfplotsset{compat=1.8}
\pgfarrowsdeclarecombine{twotriang}{twotriang}%
{stealth'}{stealth'}{stealth'}{stealth'}

\def\QED{\hskip0.1em\hfill\null\ \null\nobreak\hfill
\kern3pt\lower1.8pt\vbox{\hrule\hbox
{\vrule\kern1pt\vbox{\kern1.7pt \hbox{$\scriptstyle
QED$}\kern0.2pt}\kern1pt\vrule}\hrule}}

\newtheorem{example}{Example}[section]

\title{Vlasov's beams and multivector Grassmann Statics}
\author{Marcelo Epstein\footnote{University of Calgary, Canada.}}

\date{}
\begin{document}
\maketitle

\section{Introduction}

It was not until the early years of the 19th century that, thanks to the work of Louis Poinsot (1777-1859) \cite{poinsot}, the concept of {\it couple} became established and eventually adopted as one of the two fundamental concepts of solid mechanics. In time, forces and couples were regarded as the counterparts of rigid-body translations and rotations, whose formalization had already been achieved through the pioneering work of Giulio Giuseppe Mozzi (1730-1813). In a book published in 1763  \cite{mozzi}, he proved that every rigid-body motion can be represented as a ``twist'', that is, a combination of a translation along a line and a rotation about this line, a result often attributed to Michel Chasles (1793-1880).

From a formal point of view, forces and couples belong to a hierarchy of multi-vectors in the graded exterior algebra of an affine space. Thus, a whole discipline of Statics can be erected over this scaffolding, and more general kinds of mechanical interactions may be contemplated. The formulation can be carried out in an arbitrary finite-dimensional framework, but even the case of dimension 3 offers surprising perspectives. Moreover, when the formulation is cast in terms of the theory of screws, as conceived, among others, by Pl\"ucker \cite{plucker2}, Ball \cite{ball} and von Mises \cite{vonmises}, a sharp distinction can be drawn between an invariable part, or {\it core}, of the screw and a point-dependent field part, each of which plays an important role in the representation. It is in the interplay between these two parts that the concept of static couples of various degrees can be most properly enframed.

Following in the footsteps of our previous presentation \cite{epstein} of these ideas in the aforementioned general theoretical context, our intention here is to show how they may be implemented in a practical application. It was already remarked in \cite{epstein} that a natural materialization of the theory is to be found in the area of bodies with internal microstructure, such as Cosserat media. After careful consideration, however, it became clear that a more down-to-earth application would better illustrate the main underlying ideas of the general formulation in a more palpable fashion.

It is a fortunate circumstance that the noted Russian applied mechanics master Vasilii Zakharovich Vlasov (1906-1958), in his elegant and momentous treatise \cite{vlasov}, first published in Russian in 1940, not only anticipated some of the features of multivector statics but also created the terminology of `bimoments', which is most suited to the exterior algebra setting. Vlasov's theory of thin-walled elastic beams is essentially a structural engineering model, with all its concomitant advantages of visualization and appeal to physical intuition. Furthermore, it can be also placed within the context of Cosserat media, as proposed in \cite{epstein2, epstein1}, thus becoming a stepping stone toward a more general treatment of media with internal structure by means of exterior algebraic concepts.

\section{The algebra of multivectors}

It is ironic, but not altogether surprising in the convoluted history of science, that a fully-fledged multivector algebra, including part of its modern terminology, had been developed well before vector algebra itself.\footnote{A detailed account of the vicissitudes of the birth of vector algebra and analysis can be found in \cite{crowe}.} Guided by applications to the theory of tides and to electrodynamics, Hermann Grassmann (1809-1877) conceived a truly revolutionary {\it theory of extension}, a precursor of the modern approach to abstract algebra. Grassmann's original work \cite{grassmann} appeared in print in 1844, and received scant attention from the mathematics community of his time. It is quite remarkable that, even in our own day, Grassmann's {\it exterior algebra}, well-known and completely developed by mathematicians, has not attained universal recognition in engineering. Mechanics was indeed one of the sources of inspiration of Grassmann's mathematical work, as was also the starting point of similar mathematical ideas promulgated by Grassmann's contemporary, Adh\'emar Jean Claude Barr\'e de Saint-Venant (1797-1886).

In a vector space $V$ of dimension $n$, only two operations (vector addition and multiplication by a scalar) are defined. To have an algebra of vectors, we need to introduce an internal multiplication which is associative and distributive with respect to addition and consistent with the multiplication by scalars. Grassmann's idea was to construct an algebra not directly over $V$ but rather on an extended version of $V$ that includes scalars, vectors and multivectors (or $r$-vectors) of all orders $r$ up to and including $n$. Scalars and vectors are identified, respectively, with 0-vectors and 1-vectors. The collection of $r$-vectors for a fixed $r$ is itself a vector space, denoted by $\Lambda^r(V)$. The direct sum $\bigoplus\limits_{r=0}^n \Lambda^r(V)$ of all these vector spaces is denoted by $\Lambda(V)$. An element $\alpha$ of this vector space is, therefore, an ordered ($n+1$)-tuple of $r$-vectors, where $r$ runs from $0$ to $n$. We will refer to these elements as {\it multivector complexes}, or just {\it complexes}. Thus, a complex is of the form $\alpha=(\alpha_0, \alpha_1,...,\alpha_n)$, where $\alpha_0$ is a scalar, $\alpha_1$ is a vector, and so on. The dimension of each space $\Lambda^r(V)$ will be determined below.

The new operation is denoted by $\wedge$ and is called {\it exterior} or {\it wedge product}. This operation takes an element of $\Lambda^r(V)$ and an element from $\Lambda^s(V)$, such that $r+s \le n$, and produces and element of $\Lambda^{r+s}(V)$.\footnote{If $r+s > n$ the result vanishes automatically.} Under this operation, we obtain a so-called {\it graded algebra}. As in any algebra, the new operation satisfies the associative, distributive and consistency conditions
\begin{equation} \label{alg1}
\alpha\wedge (\beta\wedge \gamma)=(\alpha\wedge \beta)\wedge \gamma,
\end{equation}
\begin{equation} \label{alg2}
 \gamma\wedge (\alpha+\beta )=\gamma\wedge \alpha + \gamma\wedge \beta,
\end{equation}
\begin{equation} \label{alg3}
(\alpha + \beta ) \wedge \gamma=\alpha\wedge \gamma+ \beta \wedge \gamma,
\end{equation}
and
\begin{equation} \label{alg4}
a(\alpha\wedge \beta)=(a \alpha)\wedge \beta =\alpha\wedge(a\beta)\;\;\;\;\;\forall a \in {\mathbb R}.
\end{equation}
In Equations (\ref{alg1}) and (\ref{alg4}), $\alpha, \beta, \gamma$ are multivectors of any order, while in Equations (\ref{alg2}) and (\ref{alg3}) $\alpha$ and $\beta$ are of the same order. Moreover, inspired by the properties of oriented segments, areas and volumes, the following (anti-commutativity) property is assumed
\begin{equation} \label{alg5}
\alpha\wedge \beta = (-1)^{rs} \beta \wedge \alpha\;\;\;\;\;\forall \alpha \in \Lambda^r(V),\; \;\beta \in \Lambda^s(V).
\end{equation}

It is not difficult to verify that all these properties are automatically satisfied by completely skew-symmetric contravariant tensors under the operation of skew-symmetrized tensor product. By the so-called {\it universal property} of the tensor product, one can in fact identify this operation with the exterior product, and skew-symmetric contravariant $r$-tensors with $r$-vectors on $V$.

The skew-symmetric part of a (contravariant) tensor $\bf T$ of degree $r$ over the $n$-dimensional vector space $V$ can be expressed in terms of components as
\begin{equation} \label{alg6}
\left({\rm skew\;} T\right)^{i_1,...,i_r}=\frac{1}{r!}  \; \delta^{i_1,...,i_r}_{j_1,...,j_r}\;T^{j_1,...,j_r}.
\end{equation}
In this expression, the summation convention in the range $1,...,n$ is in force for diagonally repeated indices, and the {\it generalized Kronecker symbol} is defined by
\begin{equation} \label{alg7}
\delta^{i_1,...,i_r}_{j_1,...,j_r} = \left\{
\begin{matrix}
1 & {\rm if~} i_1,...,i_r\;{\text{form an even permutation of distinct}\;} {j_1,...,j_r} \\
-1 & {\rm if~} i_1,...,i_r\;{\text{form an odd permutation of distinct}\;} {j_1,...,j_r} \\
0 & \text{otherwise}
\end{matrix}
\right.
\end{equation}
\begin{example}{\rm For $r=3$ and $n=4$, the entry $\left({\rm skew\;} T\right)^{124}$ is obtained as
\begin{equation}
\left({\rm skew\;} T\right)^{124}=\frac{1}{6} \left(T^{124}-T^{142}+T^{241}-T^{214}+T^{412}-T^{421}\right).
\end{equation}
}

\hfill{$\square$}
\end{example}

A basic combinatoric exercise shows that the dimension of the space $\Lambda^r(V)$ is
\begin{equation} \label{alg9}
\dim \Lambda^r(V)=\frac{n!}{r! (n-r)!},
\end{equation}
and
\begin{equation} \label{alg10}
\dim \Lambda(V)=2^n.
\end{equation}
We remark that $r$-vectors with $r>n$ vanish automatically, since at least two indices in each possible component will have the same value, contradicting the total skew-symmetry of the corresponding tensor.

\begin{example}{\rm For $n=3$, a complex $\alpha=(\omega_0, \omega_1, \omega_2, \omega_3) \in \Lambda(V)$ can be represented in component form as
\begin{equation}
\alpha = \left((\omega), (\omega^1, \omega^2, \omega^3), (\omega^{12}, \omega^{13}, \omega^{23}), (\omega^{123})\right).
\end{equation}
}

\hfill{$\square$}
\end{example}

\begin{example}{\rm For $n=3$, we can calculate the wedge product of two 1-vectors $\bf u, v$ as
\begin{equation}
\left(u \wedge v\right)^{ij} = - \left(v \wedge u\right)^{ij}=\frac{1}{2!} \delta^{ij}_{pq} u^p v^q= \frac{1}{2} \left(u^iv^j - u^j v^i \right).
\end{equation}
Only in case we have a pre-defined dot product in $V$ can we identify this 2-vector with an ordinary vector (the cross product) in $V$.
}

\hfill{$\square$}
\end{example}

\begin{example}\label{ex:euclidean} {\rm {\bf The cross product in ${\mathbb R}^3$}: When $V={\mathbb R}^3$, we have at our disposal the standard metric tensor with components $g_{ij}=\delta_{ij}$. The alternating symbol $\varepsilon_{ijk}=\delta^{123}_{ijk}$ behaves as a Cartesian tensor, as long as we preserve the handedness of the axes. Under these conditions, we can uniquely associate to any 2-vector $\bf m$, with components $m^{ij}=-m^{ji}$, a 1-vector $\bf M$ with (covariant) components
\begin{equation} \label{cartesian1}
 M_k =  \varepsilon_{kij} m^{ij}.
\end{equation}
Its contravariant components are numerically equal to these, namely, $M^h=\delta^{hk} M_k$. Let $\bf u$ and $\bf v$ be 1-vectors and let ${\bf m}= {\bf u} \wedge {\bf v}$. We can verify directly that
\begin{equation}
{\bf M}= {\bf u} \times {\bf v},
\end{equation}
where $\times$ denotes the ordinary cross product of vectors in ${\mathbb R}^3$.}

\hfill{$\square$}
\end{example}

\section{Affine spaces}

An $n$-dimensional {\it affine space} $\mathcal A$ is a set of elements, called {\it points}, with the following property. To each pair of points $p, q \in {\mathcal A}$ a unique vector, denoted as $q-p$ or as $\overrightarrow{pq}$, in a fixed $n$-dimensional vector space $A$ is assigned such that, for all $p, q, r \in {\mathcal A}$, this assignation
\begin{enumerate}[(i)]
\item\label{prop1} is anticommutative, that is,
\begin{equation} \label{aff1}
p-q=-(q-p);
\end{equation}
\item\label{prop2} satisfies the {\it triangle rule}
\begin{equation} \label{aff2}
q-p = (q-r) + (r-p);
\end{equation}
\item\label{prop3} is unique in the sense that, fixing any {\it origin} $o \in {\mathcal A}$, for each $p \in {\mathcal A}$ there is a unique ${\bf v}_p \in A$ such that
\begin{equation} \label{aff3}
p-o = {\bf v}_p.
\end{equation}
\end{enumerate}

The last property is sometimes expressed, quite justifiably perhaps, by saying that an affine space is a vector space without an origin, or without a preferred zero element. Once this is chosen arbitrarily, there is a unique correspondence between points and vectors. The second property is nothing but the entrenching of the sum of vectors as the closure of a polygon of forces.

If we choose an origin $o \in {\mathcal A}$ and a basis ${\bf e}_i\;(i=1,...,n)$ in $A$, we can assign uniquely to each point $p \in {\mathcal A}$ the components $x^i\;(i=1,...,n)$ of the vector $p-o$ in the chosen basis. In this way, we have established an {\it affine coordinate system} in $\mathcal A$.

Given a point $p \in {\mathcal A}$, the collection of all the vectors of the form $q-p$, where $q$ runs over the whole of $\mathcal A$, is a perfect copy of the underlying vector space $A$. Conceptually, though, every such copy is a distinct entity called {\it the tangent space to $\mathcal A$ at the point $p$}, denoted by $T_p{\mathcal A}$. By Property \ref{prop3}, each base vector ${\bf e}_i$ can be mapped uniquely to a vector at each $p$, thus providing us with an induced {\it local basis} of $T_p{\mathcal A}$. A vector at $T_p{\mathcal A}$ can be declared to be {\it equivalent} or {\it equipollent} to another vector at $T_q{\mathcal A}$ if they both have the same corresponding components in the respective local bases of any affine coordinate system. The exterior algebra $\Lambda(A)$ induces corresponding exterior algebras at each tangent space $T_p{\mathcal A}$, with the associated equivalence between multivectors.

A {\it vector field} on $\mathcal A$ is a (smooth) assignation of a vector in $T_p{\mathcal A}$ at each point $p \in {\mathcal A}$. A vector field is {\it constant} if its components are constant in the local bases induced by any affine coordinate system. In some contexts a constant vector field is sometimes referred to as a {\it free vector}. Any vector in the underlying vector space $A$ can be trivially extended to a constant vector field on the whole of $\mathcal A$. Multivector fields and complex fields can be defined in the same way. A constant field of complexes is also called a {\it free complex}.

\section{Affine screws}

A {\it screw} $\phi: p \mapsto \phi(p) \in \Lambda(T_p{\mathcal A})$ on an affine space $\mathcal A$ is a special kind of complex field. It is characterized by the existence of a {\it core} $\tilde{\phi} \in \Lambda(A)$ with the property
\begin{equation} \label{scr1}
\phi(q)=\phi(p) + (p-q) \wedge \tilde{\phi}\;\;\;\;\;\;\forall p,q \in {\mathcal A}.
\end{equation}
We should remark that the core $\tilde{\phi}$ is a fixed complex in $A$. Moreover, the wedge product between a 1-vector and a complex, appearing on the right-hand side of Equation (\ref{scr1}), has not been defined. With a certain abuse of notation, however, we interpret this multiplication as producing a new complex in which every multivector of $\tilde{\phi}$ has been pre-multiplied by $p-q$. Thus, the $r$-vector entry $\tilde{\phi}_r$ is mapped to the ($r+1$)-vector $(p-q) \wedge \tilde{\phi}_r$ of the resulting complex. As a consequence of this convention, the first entry of the result is left vacant, and the last entry of $\tilde{\phi}$ exits from the result into oblivion. We will adopt the convention that the first entry (that is, the 0-vector) of the result vanishes. Under these conditions, the statement of Equation (\ref{scr1}) makes sense.

An interesting property of screws is known as {\it equiprojectivity}. It is obtained directly from the definition and from the anticommutativity of the wedge product of a $1$-vector with itself. Indeed, we obtain
\begin{equation} \label{scr2}
(q-p) \wedge \left(\phi(q)-\phi(p)\right)=0,
\end{equation}
where the right-hand side denotes the zero complex.

A screw $\phi$, as a multivector complex field, is completely determined by its core $\tilde{\phi}$ and by its value $\phi(p)$ at one point. This pair, indicated by $[\tilde{\phi}, \phi(p)]$, is called the {\it element of reduction} of the screw to the point $p$. As fields, screws can be added or combined linearly to produce new screws. The zero screw is the screw obtained from a zero core and a zero value of the field at any one point. It is not difficult to verify that the element of reduction to a point of a linear combination of screws is precisely the linear combination of the elements of reduction of these screws to the same point.

A {\it counter-screw} $\omega: p \mapsto \omega(p) \in \Lambda(T_p{\mathcal A})$ is a field of multivector complexes that abides by the rule
\begin{equation} \label{scr3}
\omega(q)=\omega(p) + \tilde{\omega} \wedge (p-q)\;\;\;\;\;\;\forall p,q \in {\mathcal A}.
\end{equation}
The fixed complex $\tilde{\omega}$ is the core of the counter-screw. Thus, the only difference between screws and counter-screws is the order of the wedge product with the core. An element of reduction of a counterscrew to a point $p \in {\mathcal A}$ is indicated by $\{{\tilde \omega},\omega(p)\}$. Counter-screws play an important role in kinematic considerations.

\section{Statics}

The very notion of screw has its origins in mechanics, as evidenced by the pioneering works of Pl\"ucker, Ball and von Mises. In essence, for the case in which the affine space is identified with the Euclidean 3-dimensional space of classical mechanics, and when the core $\tilde{\phi}$ of the screw vanishes except for the 1-vector entry (recognized as the resultant force) while the field itself $\phi(p)$ represents the total moment of all the forces with respect to $p$, the usual formulation of Statics is recovered provided that the wedge product is identified with the cross product of vectors. The total moment, which in principle is a 2-vector, can be considered as a pseudo-vector via the metric correspondence between skew-symmetric matrices and vectors in ${\mathbb R}^3$, as shown in Example \ref{ex:euclidean}. In this context, a screw is called a {\it wrench}, a terminology introduced by Ball \cite{ball} and adopted in English-speaking countries.\footnote{The French adhere to the term ``torseur'', introduced by Appell \cite{appell}. The work of von Mises \cite{vonmises} prefers the term ``Motor''.}

Although also inspired by physical applications, Grassmann\footnote{It is interesting to point out that Grassmann was also a humanist who excelled in the field of linguistics.} was careful to distinguish between what he called real and formal sciences,\footnote{Thus, the Introduction to his opus magnum \cite{grassmann} starts with the statement; ``Die oberste Theilung aller Wissenschaften ist die in reale und formale ...'', that is, ``The principal division of the sciences is into the real and the formal.''} and advocated the pursuit of the truth of the latter as well as of the former. He thus created what he called a {\it theory of forms} of which Mechanics would be an application to ``the fundamental perceptions of the sensible world.'' In this spirit, it appears that the pursuit of the formal aspects of a theory that emerged originally from a very concrete application may not be a futile undertaking.

The idea of mechanical interactions that can be represented by entities that are not merely vectors in ${\mathbb R}^3$ is hardly new. Witness in continuum mechanics the use of higher-gradient theories with their corresponding hyper-stresses of various degrees and the associated boundary tractions. In the more modest context of this paper, however, we attempt to generalize the concept of forces and couples only within the framework of pure statics in an affine space of an arbitrary number of dimension and devoid of any metric structure. Revisiting the 3-dimensional case at the end of the treatment may suggest some intriguing possibilities.

Adopting the wrench as our fundamental static entity, we may say that it manifests itself at each point $p$ by its element of reduction, namely the pair $[\tilde{\phi}, \phi(p)]$, which we call a {\it static element} at $p$. Vice versa, given such a pair of complexes at a point, and regarding the first complex as the core and the second as the field value, we can uniquely construct a wrench and evaluate it at any other point of $\mathcal A$. Given two static elements at points $p$ and $q$, we say that they are {\it equivalent} if they determine the same wrench. Since the core of a screw is unique, equivalence implies equality of the first complex of each pair. The second complex must abide by Equation (\ref{scr1}).

Static elements can be added by adding the corresponding wrenches. A system consisting of a finite number of static elements $[\tilde{\phi}_I, \phi_I(p_I)]\;(I=1,...,N)$ is in {\it equilibrium} if their sum is zero. Reducing all the static elements to a point $p$, we obtain the equilibrium equations
\begin{equation} \label{st1}
\sum\limits_{I=1}^N \tilde{\phi}_I = 0,
\end{equation}
and
\begin{equation} \label{st2}
\sum\limits_{I=1}^N  \phi_I(p_I)+ (p_I - p) \wedge \tilde{\phi}_I = 0.
\end{equation}

\section{Couple entities}
\label{sec:couples}

Let us focus our attention on the combined effect of two sstatic elements with mutually balanced cores, such as $[{\tilde \gamma}, \phi(p_1)]$ and $[-{\tilde \gamma}, \psi(p_2)]$. A straightforward calculation yields the following total wrench as evaluated at any point $p \in {\mathcal A}$
\begin{eqnarray} \label{co1}
\sigma(p)&=& \phi(p)+\psi(p) \nonumber \\
&=&\phi(p_1)+(p_1-p)\wedge {\tilde \gamma} + \psi(p_2) +(p_2-p)\wedge (-{\tilde \gamma}) \nonumber \\
&=&\phi(p_1)+ \psi(p_2)+(p_1-p_2)\wedge {\tilde \gamma}.
\end{eqnarray}
In other words, the element of reduction of the total wrench $\sigma$ at a point $p$ has a zero core and the additive contribution of the (opposite) cores of the two wrenches is independent of $p$ and equals the product $(p_1-p_2)\wedge {\tilde \gamma}$. Thus, the contribution to the total field produced by two wrenches with mutually balanced cores is a ``couple'' entity which, being constant, is a free complex. The terms couple and free are borrowed from the classical definition of force couples.

\section{Kinematics and virtual power}

We have already intimated that conventional statics finds its kinematic counterpart in rigid-body motions. Thus, for example, a classical system of forces and couples is in equilibrium if, and only if, its virtual power on any rigid-body velocity field vanishes. Having extended the notion of force to include multivectors of all degrees, we must provide an appropriate kinematic counterpart, which will be called a {\it twist}. A twist $\omega$ is a counterscrew, whose core ${\tilde \omega}$ is called the {\it angular velocity complex}. The field $\omega(p)$ generated by this core is a {\it velocity complex}.  This terminology is only suggestive, and should not be interpreted literally. Given an element of reduction $\{{\tilde \omega}, \omega(p)\}$ at a point $p \in {\mathcal A}$, the field of velocity complexes is obtained from Equation (\ref{scr3}). 

The immediate formal reason to define kinematic twists as counterscrews, rather than screws, is that it is possible to define invariantly a bilinear operation between screws and counterscrews that is reminiscent of a dot product, except that it takes values in $\Lambda^n(V)$. Recall that this space is one-dimensional, just like the space of scalars. Given a screw $\phi$ and a counterscrew $\omega$, we define their {\it pseudo-dot product} as
\begin{equation} \label{kin1}
\langle\phi\;|\;\omega\rangle= \sum\limits_{r=0}^{n}\;\left( {\tilde \phi}_r\;\wedge\; \omega_{n-r}(p)
\pm {\phi}_r(p)\;\wedge\; {\tilde \omega}_{n-r}\right) .
\end{equation}
In this equation, the plus and minus signs apply, respectively, for even and odd dimension $n$. It is not difficult to show that the result is independent of the particular point of reduction $p$ adopted.

When the screw is a wrench and the counterscrew is a twist, we call their pseudo-dot product {\it the virtual work} of the wrench over the screw. Notice that if we interpret, for example, 1-vectors as forces, and 2-vectors as moments, then their kinematic duals are, respectively, interpreted traditionally as linear and angular velocities. In our expression, these quantities correspond, respectively, to ($n-1$) and ($n-2$)-vectors in the twist. From this observation, we conclude that the entries in the core of a twist should be read `backwards', as it were.

\section{An application: Vlasov's beam theory}

Consider a beam cross section as a planar entity to which forces and couples can be applied at various points. In classical (Euler-Bernoulli or Timoshenko) beam theories, this system of forces and couples can be reduced to any point in the plane of the cross section (in particular, the centroid of the cross section) to obtain a single force-couple resultant that produces the same effect as the original system. The justification of this reduction can be found in the fact that the cross section is assumed to remain rigid during the process of deformation, In Vlasov's model, however, the cross section can undergo a special kind of {\it warping}, whereby small out-of-plane normal displacements can take place. This kind of kinematic assumption is shown by Vlasov \cite{vlasov} to be suitable for the treatment of the combined bending and torsion of beams of thin-walled open (i.e., simply connected) cross section. In particular, the planes on which the couples act play an important role in the theory in the sense that when moving a couple to a parallel plane a moment-couple needs to be taken into consideration. This `moment of a moment' is what Vlasov calls a {\it bimoment}.

\begin{figure}[h!]
\begin{center}
\begin{tikzpicture} [scale=1]
\path[fill=gray!40] (2.2,0.7)--(4.8,0.7)--(4.8,3.3)--(2.2,3.3)--cycle;
\draw[fill=white, thick] (0,0) -- (2,0)--(4.5,1)--(2.5,1)--cycle;
\draw[fill=white, thick] (1,0) -- (3.5,1)-- (3.5,3) -- (1,2.) --cycle;
\draw[fill=white, thick] (0,2) -- (2,2)--(4.5,3)--(2.5,3)--cycle;
\path[fill=gray!40] (8.2,0.7)--(10.8,0.7)--(10.8,3.3)--(8.2,3.3)--cycle;
\draw[fill=white, thick] (6.02-0.5,0.45)--(7.98-0.5,-0.45)  to [bend right=5] (10.5,1)--(8.5,1) to [bend left=5] cycle;
\draw[fill=white, thick] (7.-0.5,0)--(7+0.5,2)--(9.5,3)--(9.5,1) to [bend left=5] cycle;
\draw[fill=white, thick] (6.02+0.5,2+0.45) -- (7.98+0.5,2-0.45) to [bend left=5] (10.5,3)--(8.5,3) to [bend right=5] cycle;
\draw[->, thick]  (7,2) to [bend right=35] (8,1.7);
\draw[<-, thick]  (6,0) to [bend right=35] (7,-0.3);

\end{tikzpicture}
\end{center}
\caption{A bimoment acting on a clamped I-beam.}
\label{fig:ibeam}
\end{figure}
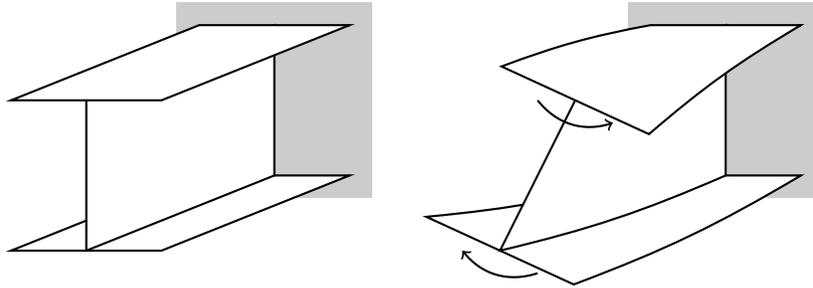

An illustration of the physical motivation behind Vlasov's conception can be gathered by considering an I-beam, as shown in Figure \ref{fig:ibeam}, loaded with two equal magnitude but opposite couples acting on the planes of the flanges. From the point of view of classical beam theory, this system is self-balanced and, therefore, it produces no stresses or deformations whatsoever.\footnote{Vlasov points out that even if the beam were to be considered as a 3-dimensional elastic body, it would be inappropriate to invoke Saint-Venant's principle without further qualification. This principle states that boundary conditions of traction can be replaced by a statically equivalent system, as long as the boundary region considered is of small extent, without affecting significantly the solution of the boundary-value problem sufficiently far from this region.} In actual fact, these two couples tend to bend the flanges in opposite directions giving rise thereby to a torsional effect.

Our objective is to show that some static and kinematic aspects of Vlasov's theory emerge naturally in the framework of multivector statics from the mere assumption that {\it the core of each wrench acting on the cross section contains both forces and moments}.

\section{Vlasov's kinematics and 0-vectors}

In Chapter 1 of Vlasov's book \cite{vlasov}, the kinematic analysis of a thin-walled beam in the small deformation regime is developed in great detail. In Vlasov's own words, when compared to the conventional Euler-Bernoulli theory, there is an additional

\begin{quote}
``part of the displacement that does not obey the law of plane sections and which arises as a result of torsion. We call this deviation from the law of plane sections the {\it sectorial warping} of the cross section. This warping is given by the {\it law of sectorial areas}.''
\end{quote}

We will presently show that the torsional warping and the law of sectorial areas are actually encoded in a single scalar quantity, namely, the first entry of the twist counter-screw. Such is the power of Grassmann's algebra.

To validate this statement, let us consider a twist whose core, given by  $\tilde \omega=({\tilde \omega}_0, {\tilde \omega}_1, {\tilde \omega}_2, {\tilde \omega}_3)$, vanishes except for the (scalar) entry ${\tilde \omega}_0$. Let us, moreover, assume that we have found a point $p_0$ in the plane of the cross section such that the corresponding field vanishes thereat. It follows from Equation (\ref{scr3}) that the field at any other point $p$ vanishes except for the 1-vector entry $\omega_1$, which is given by
\begin{equation} \label{kin1}
\omega_1 (p)=-{\tilde \omega}_0 (p-p_0).
\end{equation}
Recalling that the meaning of the 1-vector entry of the field generated in ${\mathbb E}^3$ by a twist is an angular velocity, we arrive at the conclusion that the field generated on the thin-walled cross section by our special core (with a single non-vanishing scalar entry) is a field of angular velocities proportional to the radius vector drawn from the pole $p_0$.

\begin{figure}[h!]
\begin{center}
\begin{tikzpicture} [scale=1]
\draw[ultra thick] (1,4) to[bend left=80] coordinate[pos=0.1] (A) coordinate[pos=0.35] (B) (0,0);
\filldraw (-1,2) circle (0.5mm);
\path[fill=gray!40] (A)--(B)--(-1,2)--cycle;
\draw[-stealth', thick] (-1,2) -- (A);
\draw[-stealth', thick] (-1,2) -- (B);
\draw[-stealth', thick] (A) -- (B);
\node[right] at (A) {$m$};
\node[right] at (B) {$m+\overrightarrow{ds}$};
\node[left] at (-1,2) {$p_0$};
\end{tikzpicture}
\end{center}
\caption{Vlasov's law of sectorial areas obtained from the core of a twist}
\label{fig:sectorial}
\end{figure}
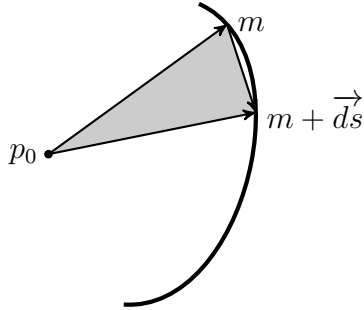

Consider two points, $m$ and $m+\overrightarrow{ds}$, along the section profile, as shown in Figure \ref{fig:sectorial}. Taking into account the metric structure of ${\mathbb E}^3$, the second point will experience an out-of-plane incremental displacement given by the cross product $-{\tilde \omega}_0 (p-p_0) \times\ \overrightarrow{ds}$, whose magnitude is, therefore, proportional to the area of the triangle shaded in the figure. We conclude that the differential out-of-plane displacement between two points, $m_1$ and $m_2$, brought about by the mere assumption of a non-vanishing scalar entry of the core of the twist, is given by the area swept by the above mentioned radius vector when going from $m_1$ to $m_2$ along the section profile. This is precisely Vlasov's law of sectorial areas.

\section{Vlasov's bimoments as 3-vectors}

A static couple entity, as discussed in Section \ref{sec:couples}, has a vanishing core and, hence, gives rise to a constant field. Specifically, a pair of equal magnitude and opposite sense $r$-vectors, applied at different points, produces a constant field of ($r+1$)-vectors. The constancy of this field has been already interpreted as a free multivector. For example, in the classical context, a pair of forces (1-vectors) produces a free force-couple (a 2-vector). These force-couples, however, by virtue of being free (that is, not having a preferred point of application), cannot possibly be a source of further multivector-couples. Vlasov understood this very well within the physical context of thin-walled beams. He states that:
\begin{quote} 
``In contrast to the moment, the bimoment is a generalized balanced force system, i.e., a force system statically equivalent to zero.'''
\end{quote}
In other words, a bimoment consists of two force-couples {\it acting at two specific points} and statically cancelling each other in the traditional sense. The component force-couples, therefore, dwell in two parallel planes.

From the point of view of multivector statics, all we need to do is to consider non-vanishing 2-vectors within the {\it core} of a wrench. Let $\tilde \phi=({\tilde \phi}_0, {\tilde \phi}_1, {\tilde \phi}_2, {\tilde \phi}_3)$ be the core of a wrench such that only the multivector ${\tilde \phi}_2$ does not vanish, and also such that at a point $p_0$ in the plane of the cross section  the corresponding field vanishes. Let ${\tilde \phi}'$ be the core of another wrench of the same kind, but such that ${\tilde \phi}'_2=-{\tilde \phi}_2$, and such that its field vanishes at a different point $p'_0$ of the plane of the cross section. It follows from Equation (\ref{scr1}) that the field generated by these two wrenches consists of a constant 3-vector given by
\begin{equation} \label{bim1}
\beta=(p'_0-p_0) \wedge  {\tilde \phi}_2.
\end{equation}
This 3-vector is precisely Vlasov's bimoment.

Not surprisingly, the bimoment (a 3-vector) is the virtual-power-dual of the scalar component of the twist (a 0-vector). This fact is stated, albeit not sufficiently emphasized, by Vlasov in the following terms:
\begin{quote}
``Starting from the notion of virtual work, we can determine the generalized longitudinal forces as the work of all the elementary longitudinal'' stresses acting on ``each of the admissible longitudinal generalized displacements that we have examined ..."
\end{quote}

\section{Conclusion}

There is much more to Vlasov's theory of thin-walled beams than what we have extracted from the Grassmann algebra approach. Indeed, we have limited ourselves to the statical and kinematical facts pertaining to a single cross section, whereas Vlasov's theory involves the concepts of stress and strain that result from the interaction between neighbouring cross sections. Our intention here, however, has been to demonstrate the potential of the multivector formalism and its surprising correspondence with the physical facts. Classical statics is the dual of rigid-body kinematics. The pertinent wrenches have cores consisting exclusively of ordinary vectorial forces. Correspondingly, the cores of the twists contain exclusively 2-vectors (skew-symmetric matrices) representing rigid-body angular velocities. We have demonstrated that, without venturing into the domain of higher-dimensional statics, but remaining in our ordinary Euclidean space, the mere augmentation of the core of the wrenches to include bound moments, and of the core of the twist to include a scalar element, has delivered the static and kinematic innovations laboriously worked out in Vlasov's thin-walled beam theory. It is to be expected that a Continuum Mechanics theory based on similar ideas will deliver known theories of media with microstructure and be the source of new models of materials. Possible candidates are certain types of liquid crystals and of biological tissues which are reinforced both by fibres and by planar elements. The reason behind these choices can be found in an important aspect of multivectors which we have not included in the present treatment, namely, the intimate connection between multivectors and linear subspaces (of all possible dimensions) of the underlying vector space.

\end{document}